\documentclass[letterpaper, 10 pt, conference, final]{ieeeconf}
\IEEEoverridecommandlockouts
\usepackage{amssymb}
\usepackage{amsmath}
\usepackage{breqn}
\usepackage{comment}
\usepackage{mathtools}
\usepackage{cite}
\usepackage{graphics}
\usepackage{algorithmic}

\usepackage{todonotes}

\numberwithin{equation}{section}

\begin{document}
\title{Verification of safety critical control policies using kernel methods}

\author{Nikolaus Vertovec
\and
Sina Ober-Bl{\"o}baum
\and
Kostas Margellos \thanks{Nikolaus Vertovec and Kostas Margellos are with the University of Oxford. Email:\{nikolaus.vertovec, kostas.margellos\}@eng.ox.ac.uk. Sina Ober-Bl{\"o}baum is with Paderborn University. Email:\{sinaober@math.uni-paderborn.de\}.}}

\maketitle

\begin{abstract}
Hamilton-Jacobi reachability methods for safety-critical control have been well studied, but the safety guarantees derived rely on the accuracy of the numerical computation. Thus, it is crucial to understand and account for any inaccuracies that occur due to uncertainty in the underlying dynamics and environment as well as the induced numerical errors. To this end, we propose a framework for modeling the error of the value function inherent in Hamilton-Jacobi reachability using a Gaussian process. The derived safety controller can be used in conjuncture with arbitrary controllers to provide a safe hybrid control law. The marginal likelihood of the Gaussian process then provides a confidence metric used to determine switches between a least restrictive controller and a safety controller. We test both the prediction as well as the correction capabilities of the presented method in a classical pursuit-evasion example.
\end{abstract}

% \fi
\IEEEpeerreviewmaketitle

\section{Introduction}
Learning-based controllers are becoming increasingly relevant for robotics applications \cite{Scholkopf2007, Gillula2012, Fisac2019, Rana2021}. As a result, a steady increase in methods that allow a system to safely learn its dynamics has been noticed. Such methods provide safety guarantees while allowing a safe exploration of the state-space \cite{Berkenkamp2017}. Commonly this is done by using a hybrid control approach, whereby a least restrictive controller is used to explore the state-space and a safety critical controller is activated whenever confidence is the safe performance of the system is lost, thus driving the system back into a safe region of the state-space. A common example of such a setup is found in collision avoidance problems, whereby the safety controller is activated to prevent the least restrictive controller from driving the system too close to an obstacle. Another common example is a pursuit-evasion problem, whereby the safety controller ensures that no evasion strategy exists and thus the pursuer is guaranteed to be able to catch the evader. For the design of the safety controller, both of these problems can be modeled as differential games \cite{Mitchell2005a}, whereby the disturbance is modeled as an adversarial player.

In \cite{Fisac2019} the authors introduce safety controllers that estimates the confidence in the underlying model and if the system is shown to start to violate the underlying safety assumption, a safety controller is activated. In a similar fashion, we will derive a confidence metric for when to activate a safety controller, but instead of the confidence metric capturing the accuracy of the system model and possible range of disturbances, we will assert our confidence in the accuracy of the safety controller itself to perform as intended. This paper focuses on the derivation of the confidence metric and will therefore not test hybrid control setups. Furthermore, as in \cite{Fisac2019}, we restrict ourselves to verifying safety controllers synthesized using Hamilton-Jacobi (HJ) reachability analysis.  

The HJ-approach has been applied to various real-world applications such as air traffic management \cite{Margellos2013}, payload optimization for multi-stage launchers \cite{Bokanowski2017}, aircraft abort landing \cite{Assellaou2016}, and in \cite{Desilles2019, Vertovec2021} we see how the approach can be used to construct the Pareto front of a multi-objective optimal control problem. All these approaches rely on the fundamental assumption that the computation of the safety controller is accurate and imposes no numerical errors. Thus, the safety controller only takes into account the variations in the dynamics and disturbances modeled prior to deployment. Yet especially high dimensional problems exponentially increase the required memory, computational cost and numerical dissipation \cite{Chen2018} and thus the numerically computed value function -- on which the optimal control computation is based -- may differ from the actual one. Furthermore, even though slight changes in the dynamics or estimated disturbances might not affect the controller, it is still necessary to recompute the underlying value function used for controller estimation. We, therefore, present a method of predicting the error of these controllers using common regression models. By sampling the error in areas of interest, we are able to predict the marginal likelihood that the computed safety controller will in fact lead to a safe system behavior.

The rest of the paper is organized into five sections. Section II contains details regarding the reachability problem, including the characterization of the backwards reachable tube (BRT) as the subzero level-set of a value function, which in turn is the unique viscosity solution of a  quasi-variational inequality. Section III discusses the basic concept of using a Gaussian process to model the error of the aforementioned value function with Section IV dedicated to a case study involving a pursuit-evasion example. Finally, Section V provides concluding remarks. 

\section{Problem Setup}
As previously discussed, a key concept in reachability analysis is to compute areas of the state-space that ensure safety as well as other specifications of the system. In the context of viability theory, such regions are known as capture basins \cite{Aubin2011} and describe the region from which a system can be driven to a given target while avoiding undesired states (such as obstacles for autonomous cars or overheating of a transformer). 

Let $x \in \mathbb{R}^n$ be the system state with Lipschitz continuous dynamics $f:\mathbb{R}^n \times U\times D \rightarrow \mathbb{R}^n$ such that 
\begin{equation}
	\dot{x} = f(x(t), u(t), d(t)), \quad t \in [-T, 0],
	\label{eqn: dynamics}
\end{equation}
The sets $U$ and $D$ are compact and denote the set of possible control and disturbance inputs. The solution of the ordinary differential equation in \eqref{eqn: dynamics} is denoted by the trajectory $\zeta: [-T, 0] \rightarrow \mathbb{R}^n$, which starts from state $x_0$ at time $-T$ and is governed by the control policy $u(\cdot): [-T, 0] \rightarrow U$ and disturbance policy $d(\cdot): [-T, 0] \rightarrow D$:
\begin{align*}
	\frac{\partial}{\partial t} \zeta(t; x_0, u(\cdot), d(\cdot)) &= 
	f(\zeta(t; x_0, u(\cdot), d(\cdot)), u(t), d(t)) \\
	\zeta(-T; x_0, u(\cdot), d(\cdot)) &=  x_0.
\end{align*}
We denote the space of optimal control and disturbance policies by $\mathcal{U}$ and $\mathcal{D}$ respectively. 

To formally discuss the aims of Hamilton-Jacobi reachability analysis, we will introduce the backward reachable tube (BRT) \cite{Chen2018}. The BRT describes the set of states from which the system can be driven into a target set $R$ within the time horizon $T$. Furthermore, the system is guaranteed to never be driven into an avoid set $A$ prior to reaching $R$. By determining the BRT, it becomes possible to formally verify that a system will behave as intended and satisfy numerous reach-avoid problems even under adversarial disturbances. 

To this end, let $R \subset \mathbb{R}^n$ be the set we would like to reach, and let us define $A \subset \mathbb{R}^n$ as the set we would like to avoid. Furthermore let $l: \mathbb{R}^n \rightarrow  \mathbb{R}$ and $h: \mathbb{R}^n \rightarrow  \mathbb{R}$ denote the Lipschitz continuous signed distance functions to $R$ and $A$, respectively. Then we can characterize the reach and avoid sets as follows:
\begin{align}
	R &= \big\{x \in \mathbb{R}^n : l(x) \leq 0 \big\} \\
	A &= \big\{x \in \mathbb{R}^n : h(x) > 0 \big\}.
	\label{eqn: reach-avoid-set}
\end{align}

Under a robust control lens, we will compute the BRT for the worst-case disturbance and thus ensure that our assumptions of reaching $R$ and avoiding $A$ hold for all possible disturbances. To this end, we reformulate our control problem as a reach-avoid differential game. The first player will determine $d(\cdot)$, while the second player will determine $u(\cdot)$.

Let us introduce the nonanticipative strategy \cite{Varaiya1967, Evans1984}, which is a function $\gamma: \mathcal{U} \rightarrow \mathcal{D}$, such that for all $t \in [-T, 0]$ and for all $u, \hat{u} \in \mathcal{U}$, if $u(\tau) = \hat{u}(\tau)$ for almost every $\tau \in [-T, t]$, then $\gamma[u](\tau) = \gamma[\hat{u}](\tau)$ for almost every $\tau \in [-T, t]$. Furthermore, we use $\Gamma$ to denote the class of nonanticipative strategies and we restrict the first player to only nonanticipative strategies. Then it is possible to describe the desired BRT as follows:
\begin{multline}
	BRT(T,R,A) =  \big\{ x \big| \exists \gamma(\cdot) \in \Gamma, \forall u(\cdot) \in \mathcal{U},  \\
	(\exists t \in [-T,0]: \zeta(t; x, u(\cdot), \gamma(\cdot)) \in R) \quad \\ 
	\& \quad (\forall \tau \in [-T, t], \zeta(\tau; x, u(\cdot), \gamma(\cdot)) \notin A) \big\}
	\label{eqn: reachavoidtube}
\end{multline}
Thus the BRT contains the set of states from which a trajectory can start at time $-T$ and reach the target set $R$ within the time horizon $T$ without passing through the set $A$. The somewhat unconventional choice of the disturbance trying to steer the system state towards the target set stems from the fact that in the case study of Section \ref{sec:casestudy}, the target set encodes the set of states for which two vehicles are in collision, so it is in favour to the disturbance to enter this set. In order to determine $BRT(T,R,A)$, we introduce a value function that captures the outcome of the differential game. 
\begin{multline}
	V(x,t) = \inf_{\gamma(\cdot) \in \Gamma} \sup_{u(\cdot) \in \mathcal{U}} \\
	\max \Big\{\min_{\tau_1 \in [-t, 0]}  l(\zeta(\tau_1; x, u(\cdot), \gamma(\cdot)),  \\
		\max_{\tau_2 \in [-t, \tau_1]} h(\zeta(\tau_2; x, u(\cdot), \gamma(\cdot)) \Big\},
\end{multline}
where $\gamma$ is trying to minimize, and $u$ is trying to maximize the maximum between the minimum over all values attained by $l$ and the maximum value attained by $h$ over the time horizon required to reach $R$. In \cite{Margellos2011} the authors show that this value function is in fact the unique continuous viscosity solution of the following quasi-variational inequality:
\begin{multline}
	\max\Big\{h(x) - V(x,t), \\
	\frac{\partial V(x,t)}{\partial t} + \sup_{u \in U} \inf_{d \in D} \frac{\partial V(x,t)}{\partial x} f(x,u,d) \Big\} = 0.
	\label{eqn: HJB}
\end{multline}
By characterizing $V(x,t)$ as a viscosity solution of a quasi-variational inequality, we now only need to optimize over possible control and disturbance inputs, rather than having an infinite dimensional optimization problem over control and disturbance policies. This makes the characterization of the BRT more ascertainable. 
Once the value function has been obtained, its subzero level-set characterizes the BRT defined in \eqref{eqn: reachavoidtube}, i.e. $BRT(T,R,A) = \big\{x \in \mathbb{R}^n \big| V(x,T) \leq 0 \big\}$. 

The optimal control policy and worst case disturbance used for trajectory reconstruction are the optimizers of 
\begin{equation}
	\sup_{u \in U} \inf_{d \in D} \frac{\partial V(x,t)}{\partial x} f(x,u,d).
	\label{eqn:Hamiltonian}
\end{equation}
Thus by saving the control and disturbance values during the computation of $V(x,t)$, we are simultaneously computing the optimal control policy.

Having computed the optimal control policy for the safety controller which we denote by $u(\cdot)$, we can plug it into a hybrid control setup. To this end, let $u_l$ denote the control policy of an arbitrary least restrictive controller. The overall control law can be defined as
\begin{equation}
    u^*(t) \coloneqq \begin{cases}
        u_l(t) & \text{if} \: V(x,t) \leq -\delta \\
        u(t)   & \text{otherwise},
    \end{cases}
    \label{eqn:hybridcontrol1}
\end{equation}
where $\delta > 0$ is a chosen tolerance. Thus if the least restrictive control policy, $u_l$, takes the system within a distance $\delta$ of the boundary of the BRT, the safety controller is activated.

\section{Using Gaussian Processes to verify a control policy}
\subsection{Error of the value function}
From the value function, it is possible to derive the optimal trajectories of both the evader and the pursuer. For each state, the value function should predict how close the pursuer is able to get to the evader. 

To illustrate the criticality the numerical error plays in safety critical control design, we consider the basic pursuit-evasion setup presented in \ref{sec:casestudy}. The aim is to determine the set of states from which a pursuer is guaranteed to be able to catch the evader. We use  $\ \widetilde{} \ $  to denote the numerical solutions. Solving the quasi-variational inequality \eqref{eqn: HJB} with terminal condition $V(x,0) = \max(l(x),h(x))$, allows us to determine the numerical value function $\widetilde{V}$ as well as the optimal control and disturbance policies $\widetilde{u}$ and $\widetilde{d}$, respectively. At time $t = -0.42$, for the relative state $[x_1=0.078, x_2=-0.51, x_3=0.20]$, we have $\widetilde{V} = 0.025 > 0$ and thus the state $[x_1=0.078, x_2=-0.51, x_3=0.20]$ does not lie in the BRT, implying it is not possible for the pursuer to catch the evader.

Let us define the value function for a given control and disturbance policy as
\begin{multline}
    V_{u, d} \coloneqq \max \Big\{\min_{\tau_1 \in [-t, 0]}  l(\zeta(\tau_1; x, u(\cdot), d(\cdot)),  \\
		\max_{\tau_2 \in [-t, \tau_1]} h(\zeta(\tau_2; x, u(\cdot), d(\cdot)) \Big\},
		\label{eqn:"true"value}
\end{multline}
where the policy $d$ is derived using a nonanticipative strategy as previously discussed. Then when applying the computed control and disturbance policy, $\widetilde{u}$ and $\widetilde{d}$, respectively,
we notice that $V_{\widetilde{u}, \widetilde{d}} = -0.01$. Thus, even the potentially sub-optimal policies $\widetilde{u}$ and $\widetilde{d}$ results in the pursuer being able to catch the evader, making the numerical characterization of the BRT inaccurate. In order to correctly predict that the state $[x_1=0.078, x_2=-0.51, x_3=0.20]$ does in fact lie within the BRT, we  need to be able to determine the difference between the true value function and the numerical value function.

Since quasi-variational inequalities, such as \eqref{eqn: HJB}, contain multivariate functions of independent variables, their solutions can be defined as the relationship between two or more variables. If the computed relationship is slightly off, we are left with an approximation of the exact solution. Thus, let us consider the error term $\epsilon$, comprised of the exact solution $V$ as well the numerical solution $\widetilde{V}$ obtained by solving \eqref{eqn: HJB}. The error can be considered as the cumulative numerical error as well as the difference to the true value function under changes in the dynamics, environment or disturbances.

\begin{equation}
	\epsilon(x, t) = \widetilde{V}(x,t) - V(x,t)
	\label{eqn:epsilon}
\end{equation}

Since $V$ is usually unknown, it is not possible to determine the error $\epsilon$ perfectly and we will instead approximate it as follows:
\begin{equation}
	\widetilde{\epsilon}(x, t) = \widetilde{V}(x,t) - V_{\widetilde{u}, \widetilde{d}},
\end{equation}
where $\widetilde{u}(\cdot)$ and $\widetilde{d}(\cdot)$ are the control and disturbance policies determined by means of \eqref{eqn:Hamiltonian} using the approximated value function $\widetilde{V}(x,t)$. For a true modeling of $\epsilon$, we would need to take $\inf_{\gamma(\cdot) \in \Gamma} \sup_{u(\cdot) \in \mathcal{U}} V_{u, \gamma}$ instead of $V_{\widetilde{u}, \widetilde{d}}$. Since the aim is to verify the performance of a control and disturbance policy, this approximation of $\epsilon$ appears to be sufficient. 

For solving \eqref{eqn: HJB}, we employ level-set methods, which have the advantage of not requiring us to parameterize the solution, yet they have the disadvantage of not guaranteeing the conservation of shape and size in an advection field. Upwinding methods greatly reduce the numerical dissipation,  yet small features will still be dominated and lost as we integrate forward. Areas of the grid that contain many such features, will thus contain larger cumulative numerical error than others. At the same time, changes to the dynamics, environment or disturbances, are likely to only affect the safety critical control policy in certain areas of the state-space.

Therefore, there is a correlation between the grid and the cumulative error. If we model the error in the context of statistics and probability theory, we are trying to find the probabilistic distribution of the error in our value function. More precisely, we are trying to determine the covariance between the variables in our value function and the error. We are thus faced with a classic regression problem. To this end we expand the notation to denote trained solutions to our regression problem with a $\ \hat{} \ $ superscript. Thus $\hat{\epsilon}$ denotes the predicted error determined by a regression model fitted to finite samples of $\widetilde{\epsilon}$ and consequently, $\hat{V}(x,t) \coloneqq \widetilde{V}(x,t) - \hat{\epsilon}(x,t)$. If $\hat{\epsilon}$ approximates $\epsilon$ (defined in \eqref{eqn:epsilon}) perfectly, $\hat{V}$ will also perfectly approximate the true solution $V$.

\subsection{Prediction using Gaussian Processes}
There are a variety of different regression models that are well studied and we briefly motivate why we opt for a Gaussian process regression in the next section. 

In general, a Gaussian process regression model is a kernel-based probabilistic model of the form
\begin{equation}
	b(x) \beta + g(x),
\end{equation}
where $b(x)$ is a set of basis functions, $\beta$ is a vector of basis function coefficients and $g(x)$ is a zero mean GP with covariance function, $k(x,\hat{x})$, i.e. $g(x) \backsim GP(0,k(x,\hat{x}))$ \cite{Rasmussen2004}.

The assumption that $g(x) \backsim GP(0,k(x,\hat{x}))$, allows us to determine a marginal likelihood. However, there is no reason to believe that the error of the value function follows a Gaussian distribution. Yet experimentation shows that, nevertheless, the error can be sufficiently approximated using a Gaussian process and the marginal likelihood offers us a framework for predicting the confidence in the estimated error. The marginal likelihood is based on the standard deviation and thus, if the standard deviation is deemed to be small, we have previously verified a similar state and the confidence in the prediction is high, while in turn a large standard deviation implies that we have little confidence in our prediction. The standard deviation can, therefore, be used in a hybrid control setup to determine when to switch between a least restrictive controller and the safety controller. We, therefore, modify the control law \eqref{eqn:hybridcontrol1} to include the standard deviation
\begin{equation}
    u^*(t) \coloneqq \begin{cases}
        u_l(t) & \text{if} \: (\hat{V}(x,t) \leq -\delta) \wedge (\sigma > \sigma_0)\\
        u(t)   & \text{otherwise},
    \end{cases}
    \label{eqn:hybridcontrol2}
\end{equation}
where $\sigma_0$ is a threshold set by the designer to determine the predefined confidence that should be maintained at all times.

\begin{comment}
The four kernels studied are the rational quadratic (RQ), squared exponential (SE), Matern 5/2, and exponential kernel. Note that since all of the covariance functions are given in a normalized form, they can be multiplied by a constant $\sigma_f^2$ in order to achieve the desired process variance.

\begin{align*}
	k_{\mathrm{RQ}}(r) = &{} (1+\frac{r^2}{2\alpha l^2})^{-\alpha}\\
	k_{\mathrm{SE}}(r) = & exp(-\frac{r^2}{2l^2}) \\
	k_{\mathrm{Matern5/2}}(r) = & (1 + \frac{\sqrt{5}r}{l})exp(-\frac{\sqrt{5}r}{l}) \\
	k_{\mathrm{SE}}(r) = & exp(-\frac{r}{l}),
\end{align*}
where $l$ is a length scale and $\alpha$ is a infinite sum of squared exponential covariance functions. The various kernels have different advantages and disadvantages depending on the application. For example, the RQ kernel allows us to model data with varying scales, yet can also lead to overly smooth behavior that might make the Matern 5/2 kernel a better fit, since it is only finitely differentiable \cite{Zhang2018}.
\end{comment}

\section{Case Study} \label{sec:casestudy}
\subsection{A Game of two identical vehicles}
In order to show how a Gaussian process can be used to determine the error of a computed value function, we will introduce a pursuit-evasion example commonly used in HJ reachability analysis.

Let us consider two identical vehicles, for example, two cars. One vehicle, the pursuer, wants to catch the other vehicle, the evader. This is done by coming within a distance $r_1$ of the evader. The evader, on the other hand, successfully outruns the pursuer if it has put a distance $r_2$ between itself and the pursuer.

To model this behavior, we will have one vehicle (the pursuer) take on the best-case behavior, i.e. it will actively try and catch the other vehicle (the evader). Meanwhile the evader will play a nonanticipative strategy and try and avoid the pursuer.

Let us introduce the relative dynamics:
\begin{align*}
	\dot{x}_1 = &{} -v_e + v_p \cos(2 \pi x_3) + d  x_2 \\
	\dot{x}_2 = & v_p \sin(2 \pi x_3) - d  x_1 \\
	\dot{x}_3 = & \frac{u-d}{2 \pi},
\end{align*}
where $x_1$ and $x_2$ represent the relative distance between the two vehicles and $x_3$ describes the relative normalized heading. $u$ and $d$ are the angular velocities of the vehicles, where $u$ will be considered our control variable, and $d$ our disturbance variable. Both the control and the disturbance will be bounded by $\overline{u}$ and $\overline{d}$. Finally, $v_e$ and $v_p$ are the constant linear velocities of the evader and pursuer. $x_1$ and $x_2$ are scaled to lie between $[-1, 1]$, while $x_3$ lies between $[0,1]$. This allows us to compare the regressions models without the influence of largely varying scales. 

Let us define the target and avoid set as 
\begin{align}
	R = &{} \{ x : ||(x_1, x_2)||_2 \leq r_1 \}, \\
	A = &   \{ x : ||(x_1, x_2)||_2 \geq r_2 \},
\end{align}
where $r_1$ is the distance between the two vehicles that needs to reached for the pursuer to have caught the evader. In turn, $r_2$ is the distance at which the evader is deemed to have outrun the pursuer. Naturally, we select $r_1 < r_2$.

To compute the maximal BRT, let us define 
\begin{equation}
	l(x) \coloneqq ||(x_1, x_2)|| - r_1,
\end{equation} 
as the signed distance to our target (reach) set, and 
\begin{equation}
	h(x) \coloneqq ||(x_1, x_2)|| - r_2,
\end{equation} 
as the signed distance to our avoid set. Next, we will define the Hamiltonian as the solution of $\sup_{u \in U} \inf_{d \in D} \frac{\partial V(x,t)}{\partial x} f(x,u,d)$. The spatial derivatives of the value function are denoted by the costate vector $p$, with elements $p_1$, $p_2$, and $p_3$, which allows us to write the Hamiltonian as
\begin{multline}
	H(x,p) =  p_1(-v_e + v_p \cos(2 \pi x_3)) + p_2 (v_p \sin(2 \pi x_3)) \\
				- \overline{d} || p_1 x_2 - p_2 x_1 - \frac{p_3}{2 \pi} || + \overline{u} \frac{p_3}{2 \pi}.
\end{multline}

\begin{figure}[h]
\centering
\includegraphics[width=0.5\textwidth]{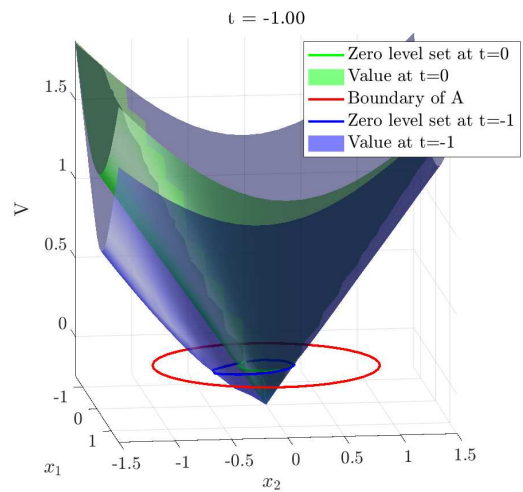}
\caption{Value function of the pursuit-evasion example in the $x_1$-$x_2$ plane sliced along $x_3 = 0.25$. The subzero level-set (within the dark blue line) indicates the BRT. The boundary of the avoid set is shown by the dark red line.}
\label{fig:value_function}
\end{figure}

Following the derivations of the previous section, this allows us to solve the resulting quasi-variational inequality, producing the final value function shown in Figure \ref{fig:value_function}. For each dimension, 21 evenly spaced grid points are used and the quasi-variational inequality \eqref{eqn: HJB} is solved until $T=1$ using a 3rd order ENO scheme together with a Lax Friedrichs Hamiltonian \cite{Osher2003}. For the computation of the value function, we set $v_e = v_p = 0.75$, and $\overline{u} = \overline{d} = 3$. Furthermore, we set $r_1 = 0.25$ and $r_2 = 1$.
%\mbox{ }

\subsection{Gaussian process error prediction}
Next, we fit a variety of regression models in order to approximate the numerical error $\widetilde{\epsilon}$.
To asses the performance of the trained models, we use 5-fold cross-validation to minimize the root-mean-square error (RMSE) of $\widetilde{\epsilon} - \hat{\epsilon}$, where $\hat{\epsilon}$ is the output of the regression model.

In Table \ref{tab:regression_comparison} the RMSE of different regression models is shown. For the comparison of the regression models, we train all models based on 1000 observations taken at random from $(x,t)$, for $x \in [-1, 1] \times [-1, 1] \times [0, 1]$ and $t \in [-1, 0]$. From the comparison of the regression models, one can see that Gaussian process regression models lead to the smallest root mean squared error.

\begin{table}[]
\centering
\begin{tabular}{ll}
\textbf{Regression Type}             & \textbf{RMSE} \\ \hline
\textbf{Linear Regression}           &               \\
Linear                               & 0.043205       \\
Interactions Linear                  & 0.041792       \\
Robust Linear                        & 0.047334        \\
Stepwise Linear                      & 0.041741       \\
\textbf{Tree}                        &               \\
Fine Tree                            & 0.031947       \\
Medium Tree                          & 0.034466       \\
Coarse Tree                          & 0.037971       \\
\textbf{SVM}                         &               \\
Linear SVM                           & 0.046966         \\
Quadratic SVM                        & 0.046893          \\
Cubic SVM                            & 0.046067         \\
Fine Gaussian SVM                    & 0.046207         \\
Medium Gaussian SVM                  & 0.045817       \\
Coarse Gaussian SVM                  & 0.046960       \\
\textbf{Ensemble}                    &               \\
Boosted Trees                        & 0.032533       \\
Bagged Trees                         & 0.039833       \\
\textbf{Gaussian Process Regression} &               \\
Squared Exponential GPR              & 0.027390      \\
Matern 5/2 GPR                       & 0.026955      \\
Exponential  GPR                     & 0.027353       \\
Rational Quadratic GPR               & 0.027029      
\end{tabular}
\caption{Comparison of commonly used regression models to approximate the numerical error of the value function.}
\label{tab:regression_comparison}
\end{table}

\begin{figure}[h]
\centering
\includegraphics[width=0.5\textwidth]{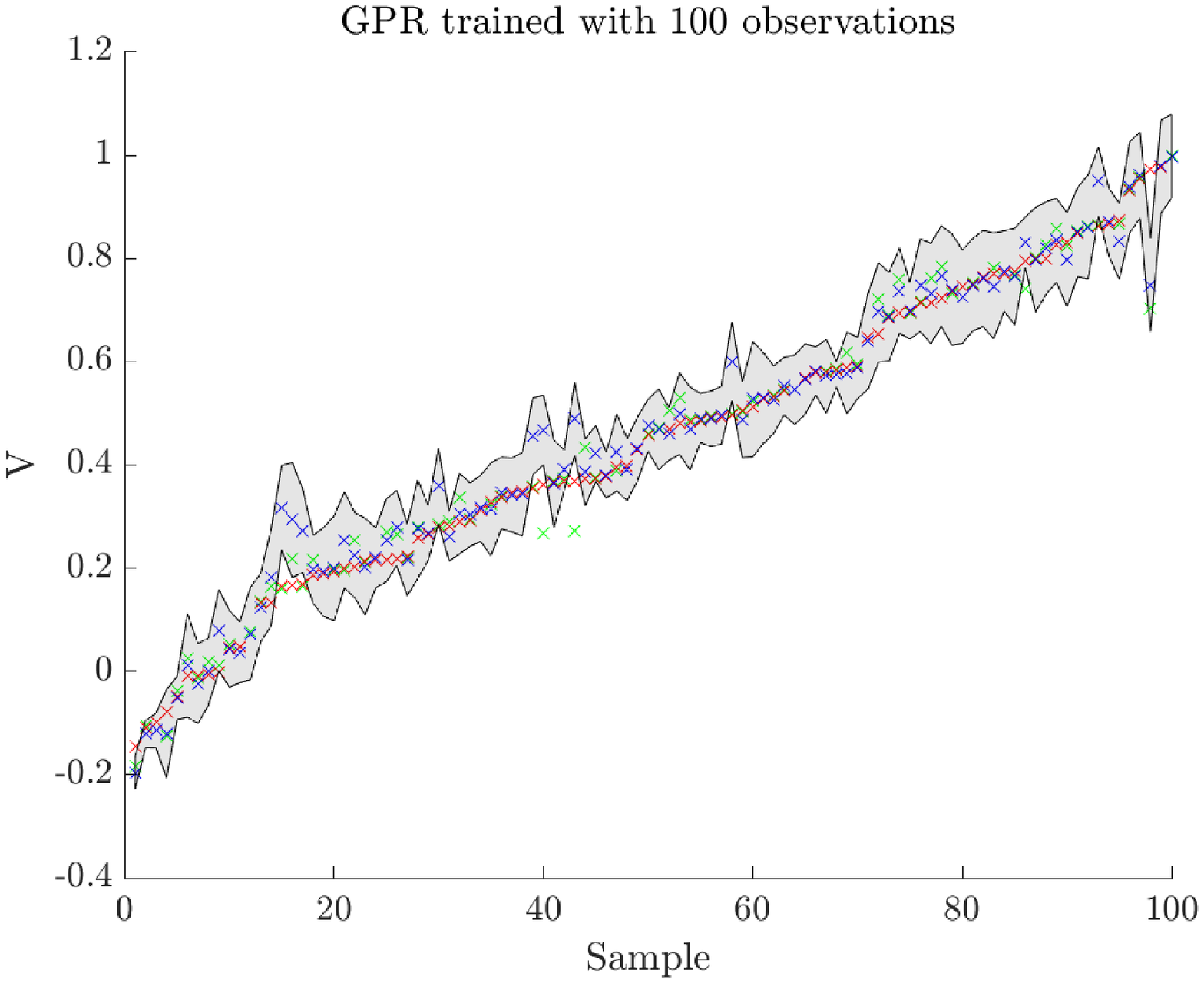}
\includegraphics[width=0.5\textwidth]{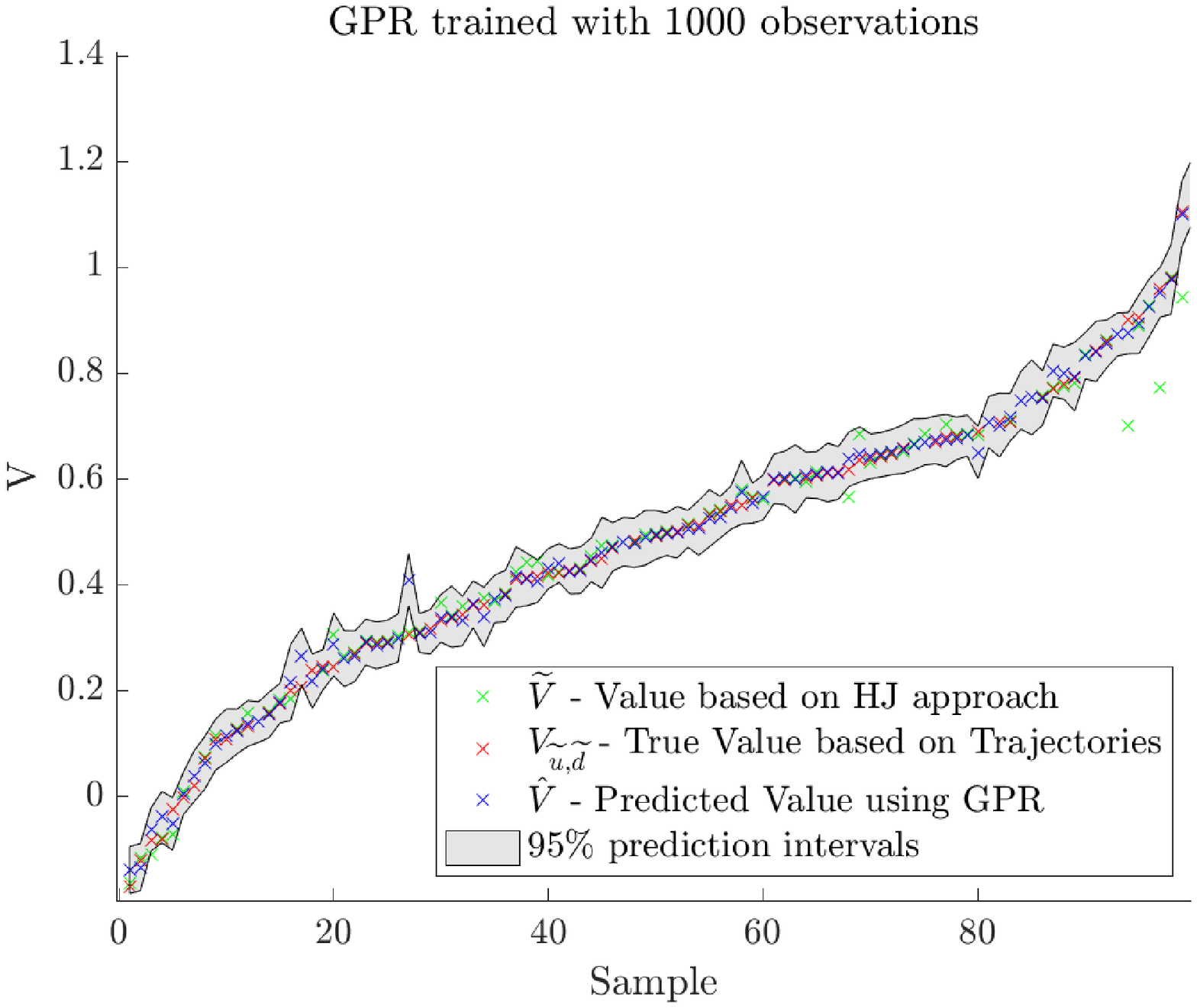}
\caption{The prediction capabilities of the Rational Quadratic GPR model trained using 100 and 1000 observation respectively}
\label{fig:error_prediction}
\end{figure}

In Figure \ref{fig:error_prediction} the prediction capability of the Rational Quadratic GPR using a constant basis function is shown. For comparison we also show the prediction capabilities of an identical regression model trained with 100, instead of 1000 observations. These two models will be referred to as the low-fidelity and the high-fidelity model, respectively. The validation set used in  Figure \ref{fig:error_prediction} consists of 100 samples taken at random from $(x,t)$, as with the training set. For each sample we look up the corresponding value from the computed value function, $\widetilde{V}$ (green points), calculate the "true" value, $V_{\widetilde{u}, \widetilde{d}}$, as in \eqref{eqn:"true"value} (red points), and compare this to the predicted value, $\hat{V}$ (blue points), obtained by subtracting the the GPR predicted error, $\hat{\epsilon}$, from the computed value function value. As can be seen, the prediction capabilities improve with a larger training set and perform at a satisfactory level at predicting the "true" value of the value function. Furthermore, even though the error does not necessarily follow a normal distribution, the $95\%$ prediction interval is able to sufficiently capture the possible range the "true" value might take. Thus using the standard deviation as a confidence metric looks promising. Based on the successful prediction capabilities, we will subsequently study how well the trained regression model can improve the control policy.

\begin{figure}[h]
\includegraphics[width=0.24\textwidth]{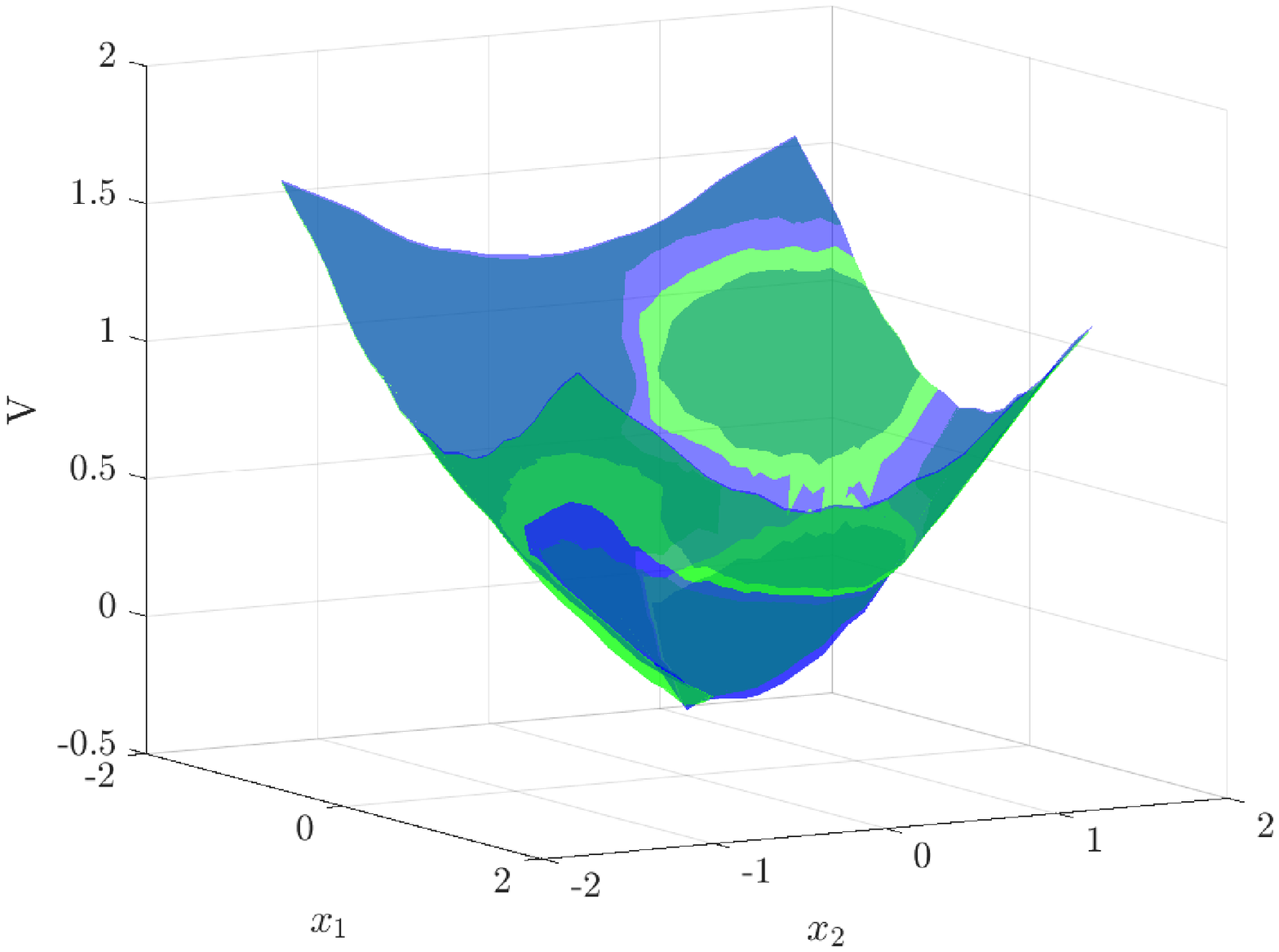}
\includegraphics[width=0.24\textwidth]{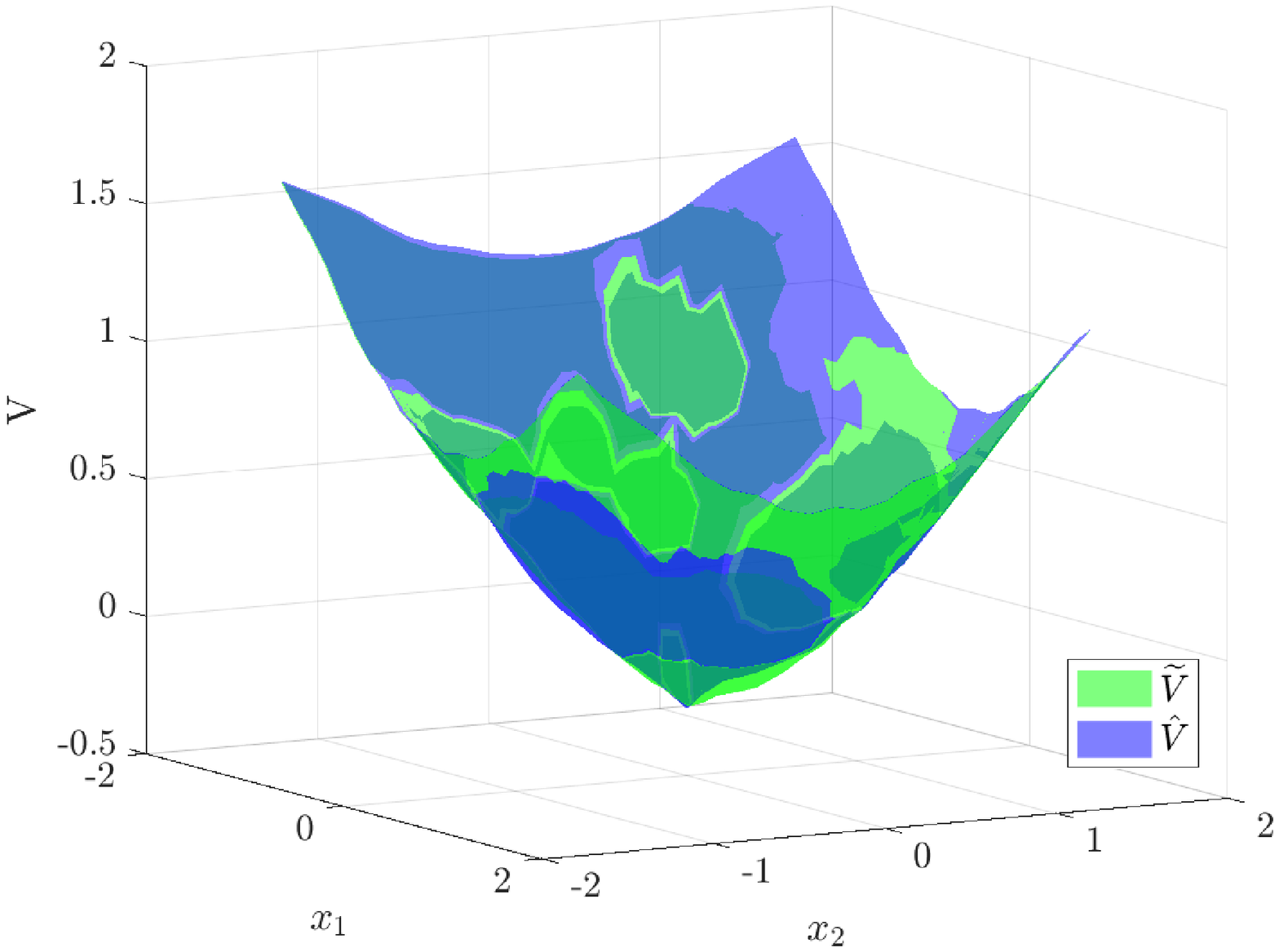}
\caption{The correction capabilities of the Rational Quadratic GPR trained using 100 (left) and 1000 (right) observation, respectively. As is expected, the regions of mismatch become more distinguished as we increase the number of observations.}
\label{fig:error_correction}
\end{figure}

Since we want to adhere to any memory constraints, we will not improve the fidelity of the value function and for every point on the grid, we simply compute the estimated error, $\hat{\epsilon}$, and subtract it from the value obtained by solving the quasi-variational inequality, $\widetilde{V}$.  The original, $\widetilde{V}$, (green) and updated value function, $\hat{V}$, (blue) using the two GPRs trained previously are shown in Figure \ref{fig:error_correction}. Using the updated value function we can modify the control and disturbance policy by recomputing the optimal minimizers/maximizers of the Hamiltonian. This results in the control and disturbance policies $\hat{u}$ and $\hat{d}$ respectively.

To evaluate the effectiveness of updating the control and disturbance policies, we calculate the RMSE of $\widetilde{V}(x,t) - V_{\widetilde{u},\widetilde{d}}$
as well as $\hat{V}(x,t) -V_{\hat{u},\hat{d}}$ using a validation set consisting of 1000 samples. As can be seen in Table \ref{tab:valfunc_comparison}, both GPR models reduce the error. Thus helping increase the accuracy of the safety critical controller significantly.

\begin{table}[]
\caption{Comparing the ability to correct the value function}
\label{tab:valfunc_comparison}
\begin{tabular}{l|l|l}
                                                       & \textbf{low-fidelity model}                & \textbf{high-fidelity model}               \\
$\widetilde{V}(x,t) - V_{\widetilde{u},\widetilde{d}}$ & $\hat{V}(x,t) - V_{\hat{u},\hat{d}}$ & $\hat{V}(x,t) - V_{\hat{u},\hat{d}}$ \\ \hline
0.0481                                                 & 0.026                                      & 0.016                                     
\end{tabular}
\end{table}
\subsection{Effects of mismatch in dynamics}
\begin{figure}[h]
\centering
\includegraphics[width=0.5\textwidth]{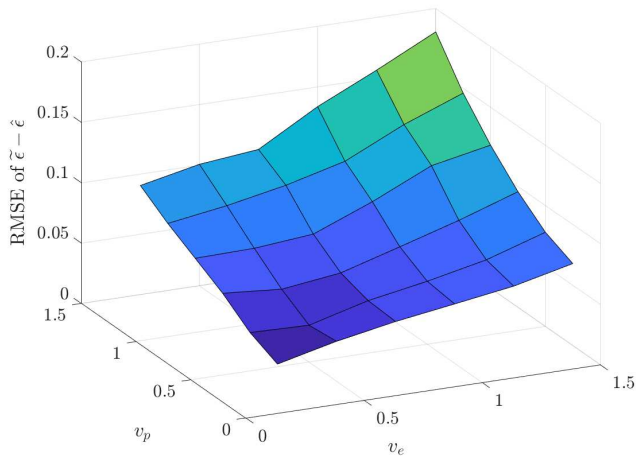}
\caption{The RMSE of a GPR model trained using 1000 samples as we vary $v_e$ and $v_p$}
\label{fig:error_prediction_mod}
\end{figure}

Next, we investigate the capabilities of detecting the error of the value function when the system dynamics of the vehicles are changed. To this end we vary $v_p$ and $v_e$ and for each variation, retrain the GPR model. Conventionally, such changes to the dynamics would require a complete recomputation of the value function. However, as can be seen in Figure \ref{fig:error_prediction_mod}, minor changes to $v_p$ and $v_e$ from the nominal values of $v_e=v_p=0.75$ do not necessary lead to a large increase in the RMSE of the GPR model ($\widetilde{\epsilon} - \hat{\epsilon}$). Naturally, as is to be expected, very large changes in the dynamics, such as doubling the speed of both the evader and pursuer will require a recomputation of the value function, since the GPR model struggles to correctly predict the error.

\section{Conclusion}
 The presented method shows that Gaussian processes can be applied to approximate the error of numerically computed backward reachable sets, even though it is not reasonable to assume that the error follows a Gaussian distribution. The standard deviation provides the means to predict the variance of the estimated error, allowing designers to determine when it is necessary to recompute the value function used for controller synthesis, or, in a hybrid control setup, when to switch between the least restrictive controller and the safety controller. Furthermore, using the prediction capabilities, we are able to update the value function, thus correcting for numerical errors and changes in the dynamics, environment or disturbances without re-solving the quasi-variational inequality associated with HJ reachability.



\bibliographystyle{./IEEEtran}
\bibliography{./IEEEabrv,./IEEEexample,./ECC22Vertovec}

\end{document}